\documentclass[reprint, superscriptaddress, amsmath, amssymb, aps]{revtex4-2}
\usepackage{multirow}
\usepackage{siunitx} % use this package module for SI units
\usepackage{amsmath}
\usepackage{amssymb}
\usepackage{braket}
\usepackage{mathtools}
\usepackage{graphicx}
\usepackage{color}
\usepackage{xcolor}
\usepackage{ulem}
\usepackage{xspace}
\usepackage{lipsum}

\usepackage[bookmarks=true,colorlinks=true,urlcolor=blue,linkcolor=blue,citecolor=blue,breaklinks]{hyperref}

\usepackage{booktabs}

\usepackage{color}
\hypersetup{
     colorlinks   = true,
     citecolor    = blue
}

\begin{document}

\title{Topological Flat Bands in Graphene Super-moiré Lattices}

\author{Mohammed M. Al Ezzi}
% \email{alezzi@u.nus.edu}
\affiliation{Centre for Advanced 2D Materials, 
National University of Singapore, 6 Science Drive 2, Singapore 117546}
\affiliation{Department of Physics, Faculty of Science, 
National University of Singapore, 2 Science Drive 3, Singapore 117542}
\author{Junxiong Hu}
\affiliation{Centre for Advanced 2D Materials, 
National University of Singapore, 6 Science Drive 2, Singapore 117546}
\affiliation{Department of Physics, Faculty of Science, 
National University of Singapore, 2 Science Drive 3, Singapore 117542}
\author{Ariando}
\affiliation{Department of Physics, Faculty of Science, 
National University of Singapore, 2 Science Drive 3, Singapore 117542}
\author{Francisco Guinea}
\affiliation{IMDEA Nanociencia, C/ Faraday 9, 28049 MADRID, Spain}
\author{Shaffique Adam}
% \email{shaffique.adam@yale-nus.edu.sg}
\affiliation{Centre for Advanced 2D Materials, 
National University of Singapore, 6 Science Drive 2, Singapore 117546}
\affiliation{Department of Physics, Faculty of Science, 
National University of Singapore, 2 Science Drive 3, Singapore 117542}
\affiliation{Department of Materials Science and Engineering, 
National University of Singapore, 9 Engineering Drive 1, 
Singapore 117575}
\affiliation{Yale-NUS College, 16 College Ave West, Singapore 138527}

\date{\today} % Leave empty to omit a date

\begin{abstract}
\normalsize
Moiré-pattern based potential engineering has become an important way to explore exotic physics in a variety of two-dimensional condensed matter systems. While these potentials have induced correlated phenomena in almost all commonly studied 2D materials, monolayer graphene has remained an exception.  We demonstrate theoretically that a single layer of graphene, when placed between two bulk boron nitride crystal substrates with the appropriate twist angles can support a robust topological ultra-flat band emerging from the second hole band.  This is one of the simplest platforms to design and exploit topological flat bands. 
\end{abstract}

\maketitle

\emph{\it Introduction}--- Since the initial isolation of graphene, these 2D sheets of carbon atoms have shown remarkable electronic properties~\cite{neto2009electronic,sarma2011electronic}.  Despite early expectations to the contrary~\cite{kotov2012electron}, the properties of monolayer graphene are mostly understood in a weakly interacting framework~\cite{tang2018role}.  There have been numerous proposals to modify the graphene's electronic properties~\cite{zhan2012engineering}, the most dramatic of these is applying an external moiré periodic potential~\cite{dos2007graphene}.  Although moiré potentials in graphene induced by a hexagonal boron nitride substrate modifies several of its electronic properties -- including the emergence of secondary Dirac points \cite{ponomarenko2013cloning} and electronic bandgaps~\cite{yankowitz2019van,yankowitz2018dynamic,jung2015origin} -- the system remains weakly correlated.

\begin{figure}[t!]
\includegraphics[width= \linewidth]{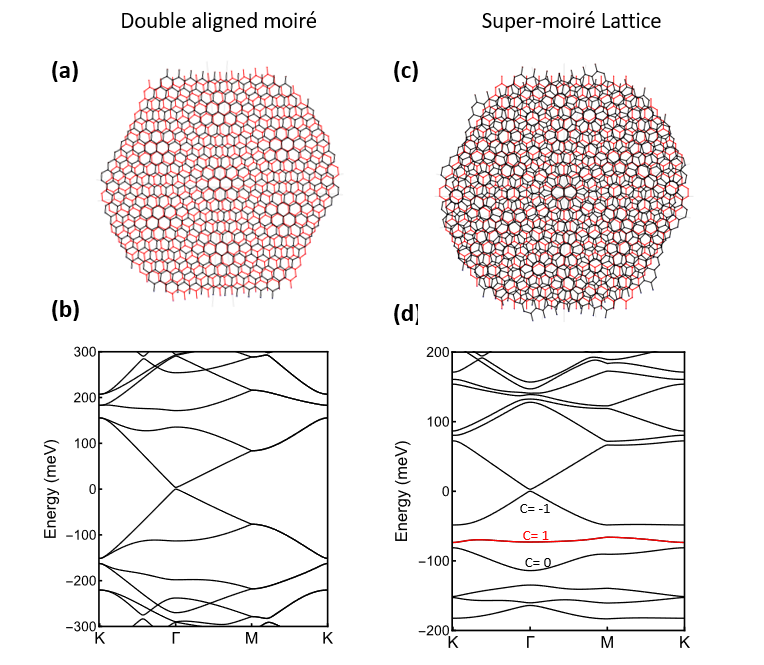}
\caption{Topological flat band in monolayer graphene with a super-moiré lattice potential. (a) Schematic  of a double aligned moiré potential where the top and bottom layers have the same twist angle  ($\theta = 0.6^\circ$).  (b) The corresponding band structure does {\it not} have flat bands.  (c) Schematic of a super-moiré lattice potential formed when monolayer graphene is encapsulated by slightly misaligned hBN substrates ($\theta_B = 0^\circ$ and $\theta_T = 0.6^\circ$).  (d) The corresponding band structure has non-zero Chern numbers for the first two hole bands with the second hole band having a bandwidth $\lesssim 10$ meV.}
\label{fig1}
\vspace{-0.15in}
\end{figure}

Twisted bilayer graphene was the first moiré-based system to show signatures of strong electron-electron interactions, a consequence of their very flat bands~\cite{bistritzer2011moire} that are often topological~\cite{zhang2019nearly}.  Experiments in such twisted moiré graphene have observed several interesting phenomena including superconductivity~\cite{cao2018unconventional}, correlated insulators~\cite{cao2018correlated}, ferromagnetism~\cite{sharpe2019emergent}, fractional Chern insulators~\cite{xie2021fractional}, etc.  Beyond twisted bilayers, correlated physics has also been seen in many families of graphene-based materials with single moiré potentials 
including twisted monolayer-bilayer~\cite{xu2021tunable,chen2021electrically,polshyn2020electrical}, twisted double bilayer~\cite{liu2020tunable}, and twisted multi-layer graphene systems~\cite{park2022robust}.

A double moiré is defined as perturbing a 2D material with two distinct moiré potentials.  This can be accomplished by encapsulating graphene with top and bottom hBN substrates~\cite{wang2019composite,wang2019new,finney2019tunable}, or using a very recently demonstrated configuration~\cite{uri2023superconductivity} where three graphene layers are twisted relative to each other with two unconstrained angles.  This often results in non-periodic or quasi-periodic potentials.  A special case of the double moiré potential is a super-moiré lattice potential, where the two moiré potentials are commensurate, but not identical.  

In this Letter, we demonstrate that a super-moiré lattice potential applied to monolayer graphene induces a flat, stable and well-separated topological band that is likely to give correlated physics similar to other graphene-based moiré systems. Since this just involves a single carbon layer encapsulated by bulk hBN cystals, our proposal is the minimal set-up required to observe correlated moiré physics in a graphene-based system and could serve as the Rosetta stone for understanding correlated phenomena in other more complicated twisted configurations.  We also find similar topological flat bands in encapsulated Bernal bilayer graphene and in Rhombohedral graphene trilayers making our proposed scheme very generic.  

Since monolayer graphene and hexagonal boron nitride (hBN) are the two most prominent two-dimensional van der Waals materials and are commonly used in a stacked combinations, one might ask if such a configuration has been previously studied.  It is well known that single-moiré superlattices lead to significant modifications to the optical and electronic properties of monolayer graphene, but the electronic bands are not flat and do not show signs of correlation physics~\cite{yankowitz2019van}.  However, some existing double-moiré devices could have unintentionally formed a commensurate super-moiré lattice potential, for example, and our proposal might have already been observed in recent experiments where about 8 percent of their double-moiré samples showed signatures of flat bands and possible correlated behaviour~\cite{sun2021correlated}.  With new experimental advances~\cite{hu2023controlled} to control the substrate alignment, it is now possible to reliably engineer such super-moiré potentials making it feasible to test these predictions.  Although there are some exotic proposals for generating flat bands in monolayer graphene e.g. by applying a periodic strain profile~\cite{mao2020evidence}, to our knowledge, inducing a topological flat band in monolayer graphene by pure moiré potentials has not previously been discussed. \\

\emph{\it Model}--- The effect of hBN on the electronic and optical properties of graphene in G/hBN heterostructures is typically included employing  either an effective periodic potential ~\cite{wallbank2013generic,san2014spontaneous,jung2017moire}, or by adopting the original continuum model for twisted bilayer graphene~\cite{dos2007graphene,bistritzer2011moire}, but replacing one graphene layer with hBN~\cite{moon2014electronic}.  Both these approaches have had considerable success in modelling single moiré potentials.  These models have also been used to study double aligned moiré potentials both with~\cite{moulsdale2022kagome} and without~\cite{sun2021correlated} a relative displacement between the two hBN stacks. To model a super-moiré lattice potential, we generalize these approaches using a Hamiltonian symbolically represented by the following 3x3 matrix:
\begin{equation}
H=\begin{bmatrix}H_{bottom-hBN} & B^{\dagger} & 0\\
B & H_{A} & T^{\dagger}\\
0 & T & H_{top-hBN}
\end{bmatrix},
\label{Eq:Hamiltonian}
\end{equation} where $H_{A}$ is the effective Hamiltonian for the ``active'' graphene subsystem (e.g. monolayer,  Bernal bilayer or rhombohedral trilayer graphene considered here) encapsulated by top and bottom hBN stacks with Hamiltonians $H_{\rm top-hBN}=H_{\rm bottom-hBN}={\rm diag}(V_{N},V_{B})$, respectively.  We use $(V_N,V_B)=(-1400,3340)$ meV for the onsite energies for nitrogen and boron atoms~\cite{moon2014electronic}. $B$ and $T$ represent the moiré coupling matrices that couple electronic states of the graphene subsystem with electronic states of the bottom and top hBN stacks respectively.

Following the usual convention (see e.g. Refs.~\cite{jung2017moire, moon2014electronic,hermann2012periodic}), we fix the lattice vectors $a_{1} =a(1,0) $ and $a_{2}  =a(\frac{1}{2},\frac{\sqrt{3}}{2})$, where $a=0.246$. The lattice vectors of the rotated hBN stacks are $a_{i}^{T/B}=MR(\theta^{T/B})a_{i}$, where $R(\theta^{T/B})$ is the rotation matrix parameterized by the amount of twist angle for top $\theta^{T}$ and bottom $\theta^{B}$ hBN stacks and $M=a_{hBN}=1.018 a$.  Stacking graphene with a given hBN substrate leads to a hexagonal single-moiré structure with lattice vectors given by $L_{i}^{T/B}=\left[1-R(\theta^{T/B})^{-1}M^{-1}\right]a_{i}$.  The super-moiré lattice occurs when the top and bottom moiré potentials are commensurate but not aligned. The super-moiré potential is hexagonal and its primitive lattice vectors given by 

\begin{equation}
    \begin{pmatrix}L_{1}\\
L_{2}
\end{pmatrix}  =\begin{pmatrix}\alpha & \beta \\
-\beta  & \alpha+\beta 
\end{pmatrix}\begin{pmatrix}L_{1}^{T}\\
L_{2}^{T}
\end{pmatrix}\\
\end{equation}

\noindent where $\alpha$ and $\beta$ are the super-moiré lattice integers. The two rotation angles $\theta^{T/B}$ and the super-moiré lattice integers for the top and bottom moire potentials are determined by satisfying the following relation 

\begin{equation}
\begin{pmatrix}\alpha & \beta \\
-\beta  & \alpha+\beta 
\end{pmatrix}\begin{pmatrix}L_{1}^{T}\\
L_{2}^{T}
\end{pmatrix} =\begin{pmatrix}\gamma & \delta \\
-\delta  & \gamma +\delta 
\end{pmatrix}\begin{pmatrix}L_{1}^{B}\\
L_{2}^{B}
\end{pmatrix}
\end{equation}

For concreteness, we align the bottom hBN stack with the graphene active subsystem $\theta^{B}=0$ and fix the top hBN stack twist angle to be $\theta^{T}\approx0.6$ degrees.  These are the smallest angles that generate a super-moiré potential (we discuss other examples at the end).  This configuration corresponds to setting $(\alpha=0, \beta=2, \gamma=1 , \delta=1 )$ in the equation above and completely specifies our model for the super-moiré periodic structure. To solve our Hamiltonian~(\ref{Eq:Hamiltonian}), we expand in the plane wave basis of the graphene subsystem and top and bottom hBN stacks and then couple them with the $B$ and $T$ matrices given by:

\begin{eqnarray}
    \{T(r),B(r)\}=w\sum_{j=0}^{3}t_{j}\left\{ e^{-iq_{j}^{\rm top}.r},e^{-iq_{j}^{\rm bottom}.r}\right\}, \notag \\
    t_{j}=\sigma_{0}+\cos(\frac{2\pi}{3}j)\sigma_{x}+\sin(\frac{2\pi}{3}j)\sigma_{y},
    \label{TB}
\end{eqnarray}
where $w= 150$ meV is the strength of the moiré potential energy and $q_j^{T/B}$ are the reciprocal lattice vectors associated with  $L_i^{T/B}$.  \\

\begin{figure}
\includegraphics[width= 1\linewidth]{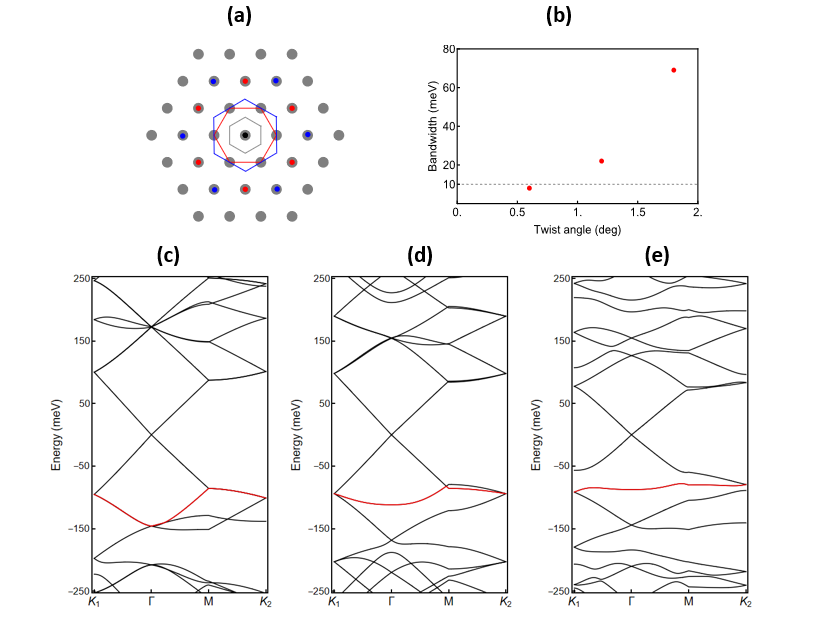}
\caption{Understanding the emergence of topological flat bands in monolayer graphene. (a) Schematic of the Brillouin zones and reciprocal lattice points of the super-moiré potential from the k-point of interest represented by the black dot at the center.  The red, blue and gray lines (dots) show the first-, second-, and super-moiré Brillouin zones (shells), respectively. (b) Bandwidth of the second hole band increases as a function of twist angle $\theta_T$ keeping $\theta_B = 0^\circ$ fixed. (c) Bandstructure of the hBN/G/hBN heterostructure including only the top moiré potential i.e. coupling only to the blue dots in panel (a).  (d) Bandstructure of the hBN/G/hBN heterostructure including only the bottom moiré potential (red dots).  (e) Bandstructure of the hBN/G/hBN with the super-moiré potential (gray dots).  Similar to our results in Fig.\ref{fig1}, the symmetry-based approach also results in an isolated and flat second hole band.  Not shown: If we use only a single moiré potential with the parameters of Ref.~\cite{wallbank2013generic}, we reproduce their results finding one mini-Dirac point in the valance band and three mini-Dirac points in the conduction band.}
\label{fig2}
\end{figure}

\emph{\it Results} --- Figure \ref{fig1} shows our main result.  The left and right panels show two different configurations and their corresponding bandstructures.  The left-panel shows a double-aligned moiré potentials (studied, for example, in Refs.~\cite{sun2021correlated,moulsdale2022kagome}).  We find the band structure for this case does not feature flat bands (in agreement with previous studies).  For the super-moiré lattice configuration considered here the electronic structure shows qualitative differences.  In particular as shown in Fig.~\ref{fig1}d, we find a flat topological band with Chern number 1. 

To further understand how this flat band emerges and examine its robustness, we apply the formalism of Ref.~\cite{wallbank2013generic} which is a general symmetry-based approach to model the effects of hBN on graphene electronic states. For example, using this approach, these authors were able to demonstrate that by parametrically tuning the coupling parameters, graphene with a single moiré potential supported mini-Dirac points at different points in the mini-Brillouin zone.  Applying this approach to the super-moiré potential, we can rewrite our Hamiltonian as: 
\begin{align}
    H&(k) = \sum_{G}\epsilon(k+G)~c_{k+G}^{\dagger}~c_{k+G} \notag \\
     &+ \sum_{G,g^{1}}\left[M(k+G,k+G+g^{1})~c_{k+G}^{\dagger}~c_{k+G+g^{1}}+h.c.\right] \notag \\
     &+\sum_{G,g^{2}}\left[M(k+G,k+G+g^{2})~c_{k+G}^{\dagger}~c_{k+G+g^{2}}+h.c.\right], 
     \label{eq2}
\end{align}
\noindent where the creation and annihilation operators for Dirac spinors are written in the expansion basis for the specific k-point.  Shown in the schematic of Fig.~\ref{fig2}a are $G$ the complete set of super-moiré reciprocal lattice vectors (gray dots), $g^{1}$ (blue dots), and $g^{2}$ (red dots) which are the first stars of the reciprocal lattice vectors of the first and second moiré potentials, respectively. For a monolayer graphene active layer,  $\epsilon(k)$ is the Dirac Hamiltonian.  For the moiré matrix elements $M(\alpha,\beta)$, we take the experimentally relevant parameter set from the analysis of Ref.~\cite{wallbank2013generic}.   

This approach allows us to develop a conceptual understanding of how the flat band emerges.  Without any super-moiré potential, we can still fold the bands into the mini-Brillouin zone.  The lattice geometry gives rise to band touching at the points of high symmetry.  We can then introduce the moiré potential as either a single moiré potential (the second line of Eq. \ref{eq2}, represented by the red dots in Fig.~\ref{fig2}a) or a super-moiré lattice potential (both the second and third lines of Eq. \ref{eq2}, represented by the gray dots).  The coupling of these terms to the planewave states of graphene can open gaps at the high symmetry points.  Since the Dirac bands (comprised of the first electron and first hole band) acquire only a tiny gap at the $\Gamma$-point~\cite{jung2015origin}, we focus instead on the second hole band. It was previously known (e.g. Refs.~\cite{wallbank2013generic, yankowitz2018dynamic, jung2017moire}) that for the second hole band, a single moiré potential opens a gap at $\Gamma$-point, while keeping the touching point at the K points (the so-called ``secondary Dirac points'').  

Zooming in close to the M point, we find that taking only the top moiré potential (blue dots), or just the bottom moiré potential (red dots) preserves the band touching (see Figs.~\ref{fig2}c and \ref{fig2}d).  However, the super-moiré potential (gray dots) opens a gap that (perturbatively) is proportional to the product of the top and bottom moiré potentials. It is this gap opening around the M point (see Fig.~\ref{fig2}e)  that makes the second hole band susceptible to isolation, while the longer real-space periodicity of the super-moiré potential acts to flatten this depinned band. 

We therefore conclude that it is the coexistence of the last two lines of Eq.\ref{eq2} that are responsible for the flat bands making it an essential property of the super-moiré potential not present in either the single moiré or aligned double moiré potentials.  This also explains the robustness of the observed phenomena with different models and parameter choices, and suggests that it would also apply to other active layers beyond monolayer graphene (see below).  While the opening of the gap at the M point can be understood from symmetry considerations, the flatness of the band depends quantitatively on the choice of parameters.  For example, in Fig.~\ref{fig2}b, we show that for realistic parameters the bandwidth is increased for larger commensurate configurations. We have checked that as expected by symmetry, the super-moiré lattice formed by the pair of twist angles ($\theta^{B}=0^\circ$, and $\theta^{T}\approx1.2^\circ$), opens gaps at the high-symmetry points. However, since the super-moiré potential is weaker, so are the the avoided crossings, and we find that the bandwidth is about twice as large. We speculate that a such a weak moiré potential with the right symmetry, but insufficiently strong to generate an isolated flat band could explain the role of lattice relaxation seen in large-scale tight-binding numerical simulations~\cite{andjelkovic2020double}. \\ 

\begin{figure}[t] 
\includegraphics[width= \linewidth]{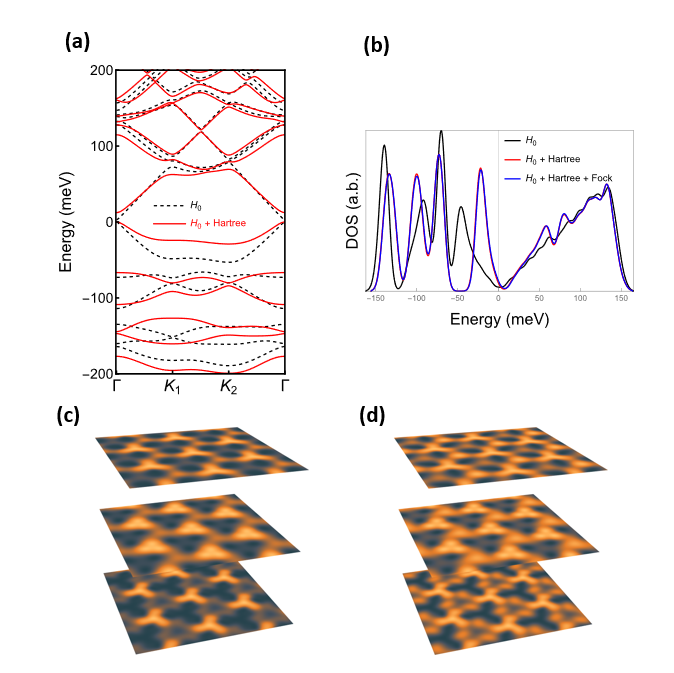}
\caption{
Effect of Hartree-Fock potentials on electronic spectrum of super-moiré hBN/G/hBN heterostructure. (a) non-interacting bandstructure without Hartree and Fock corrections (dashed black) and bandstructure for the self-consistent Hartree charge distribution correction due to filling the first hole band (solid red). (b) Density of states without Hartree-Fock corrections (black), with only Hartree corrections (blue) and with both Hartree and Fock corrections (red). The addition of Fock potential does not alter the energy spectrum of the system.  (c,d) Charge distribution of the $\Gamma$ point wavefunction without and with the effect of Hartree potential, respectively.
\texttt{}}
\label{fig3}
\end{figure}

\emph{\it Hartree-Fock Corrections}--- 
As discussed earlier, the modification to the band structure is due to the interference of moiré potentials, a purely single-particle effect. However, electrons in condensed matter systems interact with each other through the Coulomb potential, and this interaction becomes significant in flat-band systems.  We study here the robustness of the flatband to Hartree-Fock shifts of the bandstructure.  In what follows, we follow the formalism of Ref.~\cite{cea2019electronic} and we refer the reader there for technical details.  Figure \ref{fig3}a shows the band structure of monolayer with a super-moiré both with (red) and without (dashed black) the self-consistent Hartree potential.  In sharp contrast to twisted bilayer graphene~\cite{cea2019electronic}, the shape of the flat band is preserved after Hartree renormalization.  In fact, the bandwidth is slightly diminished compared to the case without interactions.  Since the second hole band remains flat and well separated from the rest of the bands, we expect that it will likely manifest experimental signatures of strongly correlated physics. 

\begin{figure}
\includegraphics[width= 1\linewidth]{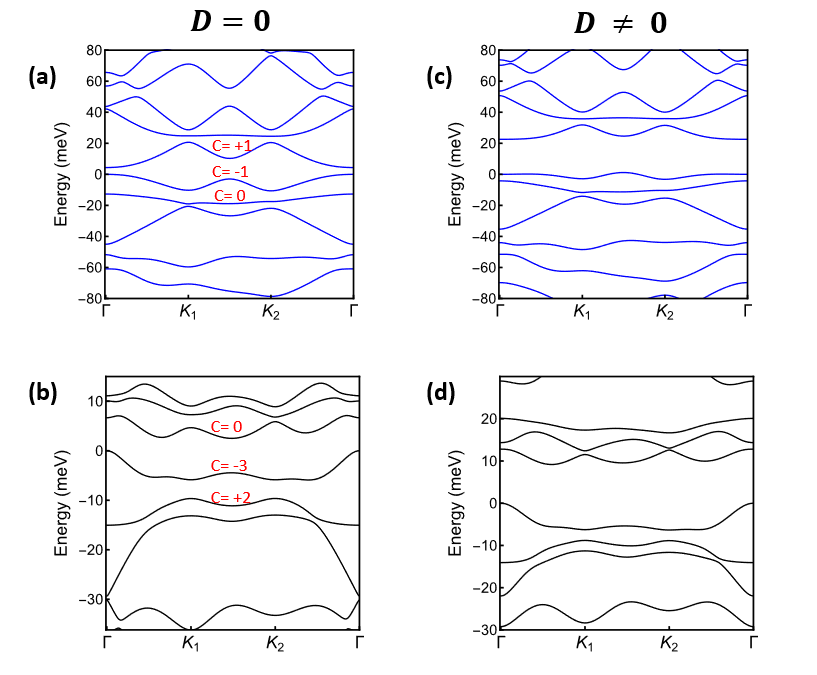}
\caption{
Super-moiré topological flat bands in bilayer and trilayer graphene, with and without electric displacement field. (a) low-energy bands of the AB bilayer graphene exposed to super-moiré potentials without external displacement field, displaying various flat bands around charge neutrality. The first electron and hole bands are topological with positive +1 or negative -1 Chern numbers respectively. (b) low-energy bands of the ABC trilayer graphene under super-moiré potential without external displacement field, where the flat bands have Chern numbers greater than 1, which is different from the case of monolayer and bilayer. (c-d) Energy bands of the AB bilayer and ABC trilayer graphene subjected to both super-moiré potential and external displacement field modelled by an electric potential energy difference of 40 meV. The indicated Chern numbers in (a-b) do not change in (c-d).  The external displacement field adjusts the width of the low-energy bands, enhancing the diversity of possible experimental phase diagrams in the bilayer and trilayer graphene super-moiré systems.
\texttt{} .}
\label{fig4}
\end{figure}

To further investigate the effect of the Hartree and Fock terms, we have calculated the density of states in three different scenarios. The results, which are illustrated in Fig.~\ref{fig3}b, show the density of states without interactions (black trace), with only the Hartree potential (red  trace), and with both the Hartree and Fock potentials (blue trace).  The blue and red traces are almost identical implying that the effect of the Fock potential is negligible in this system consistent with results on twisted bilayer graphene~\cite{cea2019electronic}.  We also show how the electron-electron interaction redistributes the charge density in real space by plotting the the modulus  of the wavefunction of the $\Gamma$-point both with and without the Hartree potential.  \\

\emph{\it Generalization to bilayer and trilayers}--- Having discovered that a super-moiré potential significantly alters the electronic properties of monolayer graphene in unexpected ways, we test whether a super-moiré has similar effects on other graphene-based systems.  We study the effect of super-moiré potentials on two other commonly used graphene systems: AB bilayer and ABC trilayer (Figure \ref{fig4}).  Remarkably, we find that the super-moiré potentials lead to the formation of topological flat bands in both bilayer graphene (where both the lowest energy electron and hole bands have Chern numbers of $\pm 1$) and in trilayer graphene.  The trilayer case is even more interesting since it supports larger Chern numbers.  Flat bands with high Chern numbers are believed to support high-temperature superconductivity \cite{peotta2015superfluidity}.  For bilayers and trilayers, the bands can be modified with an experimentally tunable displacement fields, and we show that with this additional degree of freedom, one can achieve very flat topological bands well separated from other bands. \\  

\emph{\it Conclusion}---    Our findings suggest that precise control of the alignment of the top and bottom h-BN substrates can result in a new platform for correlated physics.  In this context, we note that recent advancements in controlling hBN encapsulation of graphene with substantial yield~\cite{hu2023controlled} opens up this possibility.  We therefore anticipate experimental results along these lines in the near future.  We have demonstrated here that the interference of the two moiré potentials opens a gap in the mini Dirac points resulting in flat topological bands in untwisted 2D materials encapsulated by bulk misaligned substrates.  We speculate that applying these ideas to twisted materials like twisted bilayer graphene, twisted double-bilayer, or twisted monolayer-bilayer graphene, could result in further flattening of already flat bands or ``super-flat" bands.  We leave these speculations for a future study.

%\section*{Acknowledgements}
It is a pleasure to thank Darryl Foo, Eugene Mele, and Giovanni Vignale for helpful discussions.  We acknowledge the financial support of Singapore National Research Foundation Investigator Award (NRF-NRFI06-2020-0003) and the Ministry of Education AcRF Tier 2 grant MOE-T2EP50220-0016.  FG acknowledges support from the European Commission, within the Graphene Flagship, Core 3, grant number 881603 and from grants SEV-2016-0686 and SprQuMat (Ministerio de Ciencia e Innovación, Spain) and NMAT2D (Comunidad de Madrid, Spain).

\bibliographystyle{unsrt}
\bibliography{Bibliography}

\end{document}